\documentclass[useAMS,usenatbib,usegraphicx]{mn2e}

\usepackage{times}
\usepackage{multicol}

\def\gs{\mathrel{\raise0.35ex\hbox{$\scriptstyle >$}\kern-0.6em
\lower0.40ex\hbox{{$\scriptstyle \sim$}}}}
\def\ls{\mathrel{\raise0.35ex\hbox{$\scriptstyle <$}\kern-0.6em
\lower0.40ex\hbox{{$\scriptstyle \sim$}}}}

\title[High-resolution radio observations of SMGs]
      {High-resolution radio observations of submillimetre galaxies}
      \author[Biggs \& Ivison]
             {A.\,D.~Biggs$^{1}$\thanks{E-mail: adb@roe.ac.uk} and
              R.\,J.~Ivison$^{1,2}$\\
      $^1$UK Astronomy Technology Centre, Royal Observatory, Blackford Hill,
          Edinburgh EH9 3HJ\\
      $^2$Institute for Astronomy, University of Edinburgh, Blackford Hill,
          Edinburgh EH9 3HJ}

\date{Accepted 2007 December 17.
      Received 2007 December 14; in original form 2007 September 11}

\pagerange{000--000}

\begin{document}

\maketitle

\begin{abstract}
We have produced sensitive, high-resolution radio maps of 12
submillimetre (submm) galaxies (SMGs) in the Lockman Hole using
combined Multi-Element Radio-Linked Interferometer Network (MERLIN)
and Very Large Array (VLA) data at a frequency of
1.4\,GHz. Integrating for 350\,hr yielded an r.m.s.\ noise of
6.0\,$\mu$Jy\,beam$^{-1}$ and a resolution of 0.2--0.5\,arcsec. For
the first time, wide-field data from the two arrays have been combined
in the ($u,v$) plane and the bandwidth smearing response of the VLA
data has been removed. All of the SMGs are detected in our maps as
well as sources comprising a non-submm luminous control sample. We
find evidence that SMGs are more extended than the general $\mu$Jy
radio population and that therefore, unlike in local ultraluminous
infrared galaxies (ULIRGs), the starburst component of the radio
emission is extended and not confined to the galactic nucleus. For the
eight sources with redshifts we measure linear sizes between 1 and
8\,kpc with a median of 5\,kpc. Therefore, they are in general larger
than local ULIRGs which may support an early-stage merger scenario for
the starburst trigger.  X-rays betray active galactic nuclei (AGN) in
six of the 33 sources in the combined sample. All but one of these are
in the control sample, suggesting a lower incidence of AGN amongst the
submm-luminous galaxies which is, in turn, consistent with increased
X-ray absorption in these dust-obscured starbursts. Only one of our
sources is resolved into multiple, distinct components with our
high-resolution data. Finally, compared to a previous study of faint
radio sources in the GOODS-N field we find systematically smaller
source sizes and no evidence for a tail extending to
$\sim$4\,arcsec. Possible reasons for this are discussed.
\end{abstract}

\begin{keywords}
   galaxies: starburst
-- galaxies: formation
-- radio continuum: galaxies
-- techniques: interferometric
-- instrumentation: interferometers
-- submillimetre
\end{keywords}

\section{Introduction}

\begin{table*}
\begin{center}
\caption{Lockman Hole SMGs included in our sample. Positions and
1.4-GHz flux densities are taken from the VLA catalogue of
\citet{biggs06}. Redshifts are determined spectroscopically and are
taken from either \citet{chapman05} or \citet{ivison05,ivison07}.}
\begin{tabular}{ccccccc} \\ \hline
ID & SHADES ID   & MAMBO ID    & Bolocam ID & Radio position (J2000) & $S_{1.4}$ ($\mu$Jy) & $z$ \\ \hline
SMG01 & LOCK850.01  & LH 1200.005 & 1100.014    & 10:52:01.250 +57:24:45.76 & 73  & 2.148 \\
SMG02 & LOCK850.04  & LH 1200.003 &             & 10:52:04.226 +57:26:55.46 & 66  & 1.48 \\
SMG03 & LOCK850.09  & LH 1200.042 &             & 10:52:15.634 +57:25:04.27 & 58  & 1.853 \\
SMG04 & LOCK850.16  & LH 1200.096 &             & 10:51:51.690 +57:26:36.09 & 110 & 1.147 \\
SMG05 & LOCK850.17  & LH 1200.011 &             & 10:51:58.018 +57:18:00.26 & 93  & 2.239 \\
SMG06 & LOCK850.30  &             &             & 10:52:07.489 +57:19:04.01 & 246 & 2.689 \\
SMG07 & LOCK850.33  & LH 1200.012 &             & 10:51:55.470 +57:23:12.75 & 54  & 2.686 \\
SMG08 & LOCK850.71  &             &             & 10:52:19.086 +57:18:57.87 & 98          \\
SMG09 & LOCK850.87  &             &             & 10:51:53.362 +57:17:30.06 & 93          \\
SMG10 &             & LH 1200.008 &             & 10:51:41.431 +57:19:51.90 & 295 & 1.212 \\
SMG11 &             &             & 1100.003$a$ & 10:52:13.376 +57:16:05.41 & 249         \\
SMG12 &             &             & 1100.003$b$ & 10:52:12.256 +57:15:49.59 & 85          \\ \hline
\end{tabular}
\label{smgtab}
\end{center}
\end{table*}

SMGs form a class of object that was unknown before the advent of
modern bolometer arrays -- starting with the Submillimetre Common User
Bolometer Array \citep[SCUBA;][]{holland99} -- in the late 1990s
\citep*{smail97}.  Effective follow-up of these sources has been
particularly difficult due to the large positional uncertainty of the
submm detections. As a result, most of our understanding of the
properties of these sources has been made possible by deep, wide-field
radio imaging, which exploits the radio/far-infrared (far-IR)
correlation to produce positions with sub-arcsecond accuracy
\citep[e.g.][]{ivison02,ivison07}. It is now known that SMGs lie at
high redshift \citep{chapman03,chapman05} and it is believed that
their large far-IR luminosities are due to the ultraviolet (UV) light
produced by intense star formation ($\ge$1000\,M$_{\sun}$\,yr$^{-1}$)
being reprocessed by dust; as such they are major contributors to the
star-formation history of the Universe \citep*{hughes98}.

Although the emission from SMGs is believed to originate predominantly
from a starburst, there is likely to be a contribution from an AGN in
some sources. Signatures of AGN include X-ray emission
\citep{alexander03,alexander05}, compact or jet/lobe radio emission
\citep{chapman04,muxlow05,garrett01}, unusual radio spectral indices
as well as certain line characteristics seen in optical
\citep{swinbank04} and mid-IR spectroscopy
\citep{menendez07,lutz05}. Star formation and AGN activity are
believed to be closely linked due to the well-established correlation
between black hole and bulge mass in nearby elliptical galaxies
\citep{magorrian98,marconi03,haring04} and, furthermore, feedback
mechanisms have been proposed whereby star formation is regulated by
the AGN \citep*[e.g.][]{silk98,dimatteo05}. \citet{page04} have
suggested an evolutionary sequence within which the transition from
X-ray absorbed to unabsorbed AGN represents the termination of galaxy
growth.

In order to investigate the origin of the radio emission in SMGs, we
have obtained high-resolution imaging of the Lockman Hole with MERLIN,
a grouping of seven telescopes across the United Kingdom. The maximum
distance between telescopes, 215\,km, corresponds to an angular
resolution of $\sim$200\,milliarcsec (mas) at a frequency of 1.4\,GHz,
much higher than that obtained with the VLA at the same frequency. The
Lockman Hole is notable for its very low Galactic H\,{\sc i} column
density \citep*{lockman86} and has been targeted by a number of
different instruments at different wavelengths, including {\em
XMM-Newton} \citep{hasinger01,mainieri02,brunner07}, {\em Spitzer}
\citep{huang04,egami04} and the VLA
\citep{deruiter97,ciliegi00,ivison02}.

In this paper we combine our new MERLIN data with existing VLA data in
order to increase the sensitivity to extended emission that would
otherwise be resolved out or poorly constrained due to the lack of
short spacings, whilst still retaining significantly higher resolution
than can be obtained with the VLA alone.  The paper is structured as
follows. In the next section we describe the sample of SMGs as well as
a control sample with which the structures of the SMGs can be
compared. In \S\ref{obs} we describe the observations and the
data analysis, including a new technique for combining radio
interferometric wide-field data in the Fourier plane, and present the
maps. Discussion of our results and conclusions follow in
\S\ref{discussion} and \ref{conclusions}.

\section{Sample selection}
\label{samplesec}

\subsection{SMGs}

In the submm waveband, the Lockman Hole has been targeted in several
overlapping surveys: at 850\,$\mu$m \citep{scott02,coppin06} with
SCUBA on the James Clerk Maxwell Telescope (JCMT), at 1200\,$\mu$m
with the Max-Planck Millimeter Bolometer array (MAMBO) on the Institut
de Radio Astronomie Millim\'etrique's (IRAM's) 30-m telescope
\citep{greve04} and at 350 and 1100\,$\mu$m with SHARC\,{\sc ii}
\citep{laurent06} and Bolocam \citep{laurent05}, respectively, at the
Caltech Submillimeter Observatory (CSO). The Lockman Hole portion of
the SCUBA Half-Degree Extragalactic Survey
\cite[SHADES;][]{mortier05,coppin06} found the largest number of
sources: 60 SMGs at 850\,$\mu$m over an area of approximately
360\,arcmin$^2$. This included data from the 8-mJy Survey
\citep{scott02,fox02}. Combining source lists from several surveys has
been used as a check on the robustness of individual sources
\citep{ivison05}.

Our sample was compiled from the above references and comprises all
SMGs within 6\,arcmin of the MERLIN and VLA pointing centre
($\mathrm{10^h\,52^m\,08\fs82, +57^{\circ}\,21^{\prime}\,33\farcs8}$,
J2000) that have a peak radio flux density (as measured with the VLA)
in excess of 50\,$\mu$Jy. The radial cutoff is required to ensure that
the radio data are minimally effected by bandwidth and time smearing
(\S\ref{smearsec}) and primary beam effects. The primary beam for
unequal apertures has been investigated by \citet{strom04} and for
baselines with the most disparate telescope sizes (76\,m--25\,m) is
equal to 15.6\,arcmin ({\sc fwhm}), significantly in excess of our
12-arcmin diameter field.

With these constraints we obtain a sample of 12 SMGs which we list in
Table~\ref{smgtab}, nine of which were detected by SHADES. A number of
these sources are also detected by MAMBO and or Bolocam/SHARC\,{\sc
ii}.  The three remaining sources were only detected by either MAMBO
or Bolocam/SHARC\,{\sc ii}; the radio identification (ID) is
unambiguous in each case. For LH\,1200.008 \citep{greve04,ivison05},
there is only a single, relatively bright radio source located near
the submm position. The remaining SMG, 1100.003 \citep{laurent05}, was
originally detected by Bolocam and shown by later observations with
SHARC\, {\sc ii} \citep{laurent06} to be a blend of two sources. Both
of these have a radio counterpart; we refer to them as 1100.003a and
1100.003b respectively, where 1100.003a is much the brighter of the
two in the radio.  Flux densities and positions for all sources are
taken from \citet{biggs06} and spectroscopic redshifts (available for
seven of the SMGs) from \citet{chapman05} or
\citet{ivison05,ivison07}.

\subsection{Control sample}

As well as the SMG sample, we have defined a control sample with which
the properties of the SMGs can be contrasted. This sample is made up
of the remaining VLA detections within the 6-arcmin radius area that
have peak brightnesses greater than 50\,$\mu$Jy\,beam$^{-1}$ i.e.\ it
is defined in an identical fashion to the SMG sample. Of these, we
exclude one obviously resolved source which is much brighter than the
others, its total flux density exceeding 1\,mJy. This gives a control
sample of 21 sources.

\begin{figure*}
\begin{center}
\includegraphics[scale=0.8]{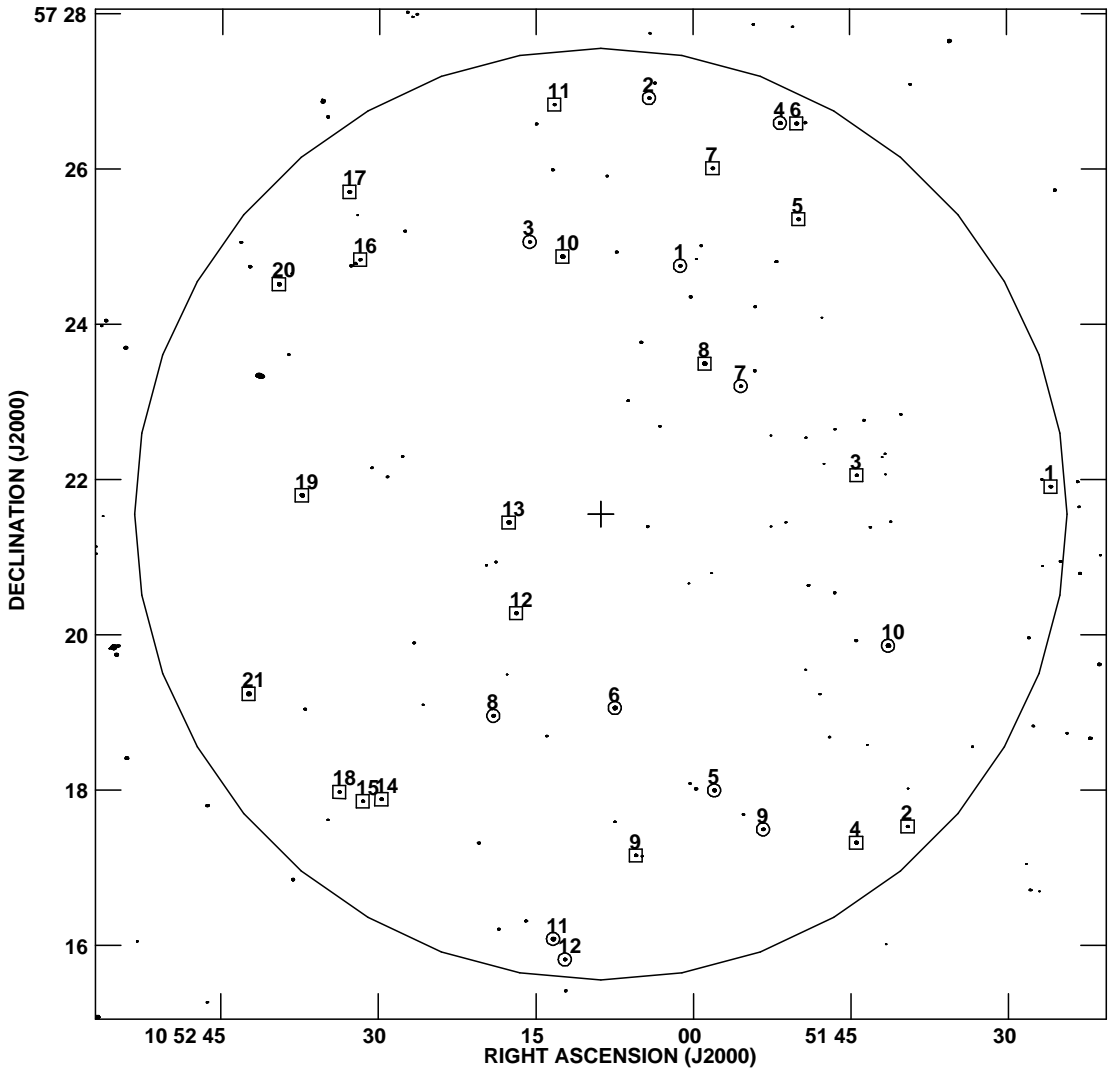}
\caption{VLA 1.4-GHz map of the Lockman Hole, showing the positions of
the SMGs (circles) and the control sample (squares). Also marked is
the 6-arcmin radial cutoff used in defining our samples and the centre
of the field.}
\label{lockmap}
\end{center}
\end{figure*}

\begin{table}
\begin{center}
\caption{Radio sources with peak flux densities greater than
50\,$\mu$Jy\,beam$^{-1}$ which lie within 6\,arcmin of the phase centre
and which do not have submm detections. These constitute our control
sample. Flux densities and positions are taken from the VLA catalogue
of \citet{biggs06}.}
\begin{tabular}{ccc} \\ \hline
ID  & Radio position (J2000) & $S_{1.4}$ ($\mu$Jy) \\ \hline
C01 & 10:51:25.892 +57:21:54.61 & 78 \\
C02 & 10:51:39.588 +57:17:32.06 & 58 \\
C03 & 10:51:44.431 +57:22:03.74 & 54 \\
C04 & 10:51:44.492 +57:17:19.66 & 137 \\
C05 & 10:51:49.939 +57:25:21.73 & 83 \\
C06 & 10:51:50.113 +57:26:35.73 & 115 \\
C07 & 10:51:58.150 +57:26:01.13 & 62 \\
C08 & 10:51:58.919 +57:23:30.13 & 224 \\
C09 & 10:52:05.491 +57:17:09.99 & 85 \\
C10 & 10:52:12.493 +57:24:52.91 & 222 \\
C11 & 10:52:13.283 +57:26:50.36 & 52 \\
C12 & 10:52:16.903 +57:20:17.24 & 108 \\
C13 & 10:52:17.615 +57:21:27.37 & 160 \\
C14 & 10:52:29.740 +57:17:53.31 & 89 \\
C15 & 10:52:31.522 +57:17:51.67 & 56 \\
C16 & 10:52:31.849 +57:24:50.39 & 55 \\
C17 & 10:52:32.877 +57:25:42.74 & 115 \\
C18 & 10:52:33.748 +57:17:58.85 & 62 \\
C19 & 10:52:37.370 +57:21:48.15 & 155 \\
C20 & 10:52:39.600 +57:24:31.37 & 143 \\
C21 & 10:52:42.418 +57:19:14.59 & 253 \\ \hline
\end{tabular}
\label{controltab}
\end{center}
\end{table}

\begin{figure*}
\begin{center}
\includegraphics[scale=2]{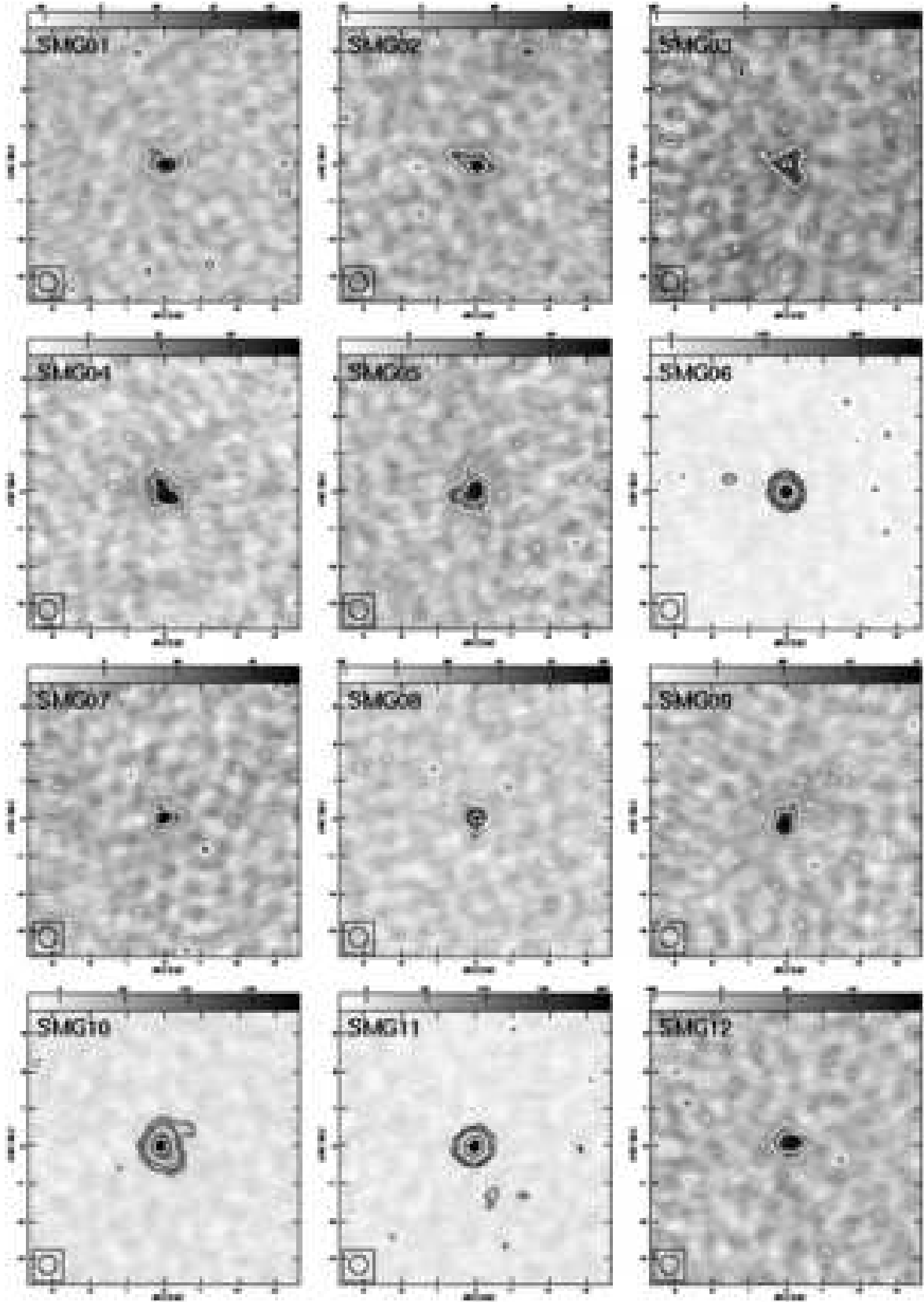}
\caption{Stokes $I$ maps of the 12 SMGs in our Lockman Hole
sample. The beam is shown in the bottom-left corner and has dimensions
of $521 \times 481$\,mas$^2$ at a position angle of $23\fdg6$.  The
greyscale has units of $\mu$Jy\,beam$^{-1}$ and contours are drawn at
$-1$, 1, 2, 4, 8, etc.\ times the 3\,$\sigma$ noise (3 $\times$
6.0\,$\mu$Jy\,beam$^{-1}$).}
\label{smgmaps}
\end{center}
\end{figure*}

\begin{figure*}
\begin{center}
\includegraphics[scale=2]{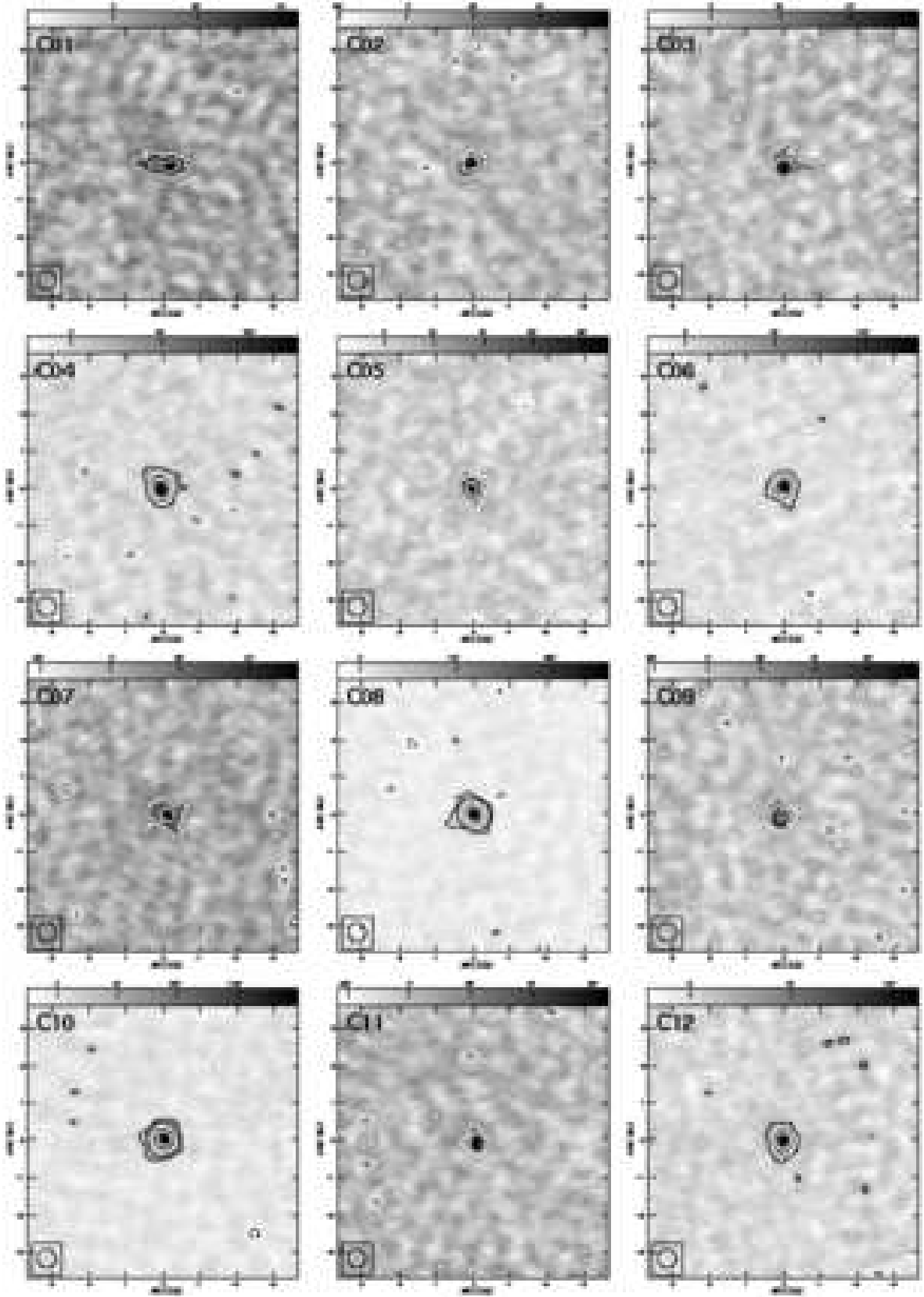}
\caption{Stokes $I$ maps of the sources in the control
sample. Contours, beamsizes and greyscales are the same as for
Fig.~\ref{smgmaps}.}
\label{controlmaps}
\end{center}
\end{figure*}

\begin{figure*}
\begin{center}
\includegraphics[scale=2]{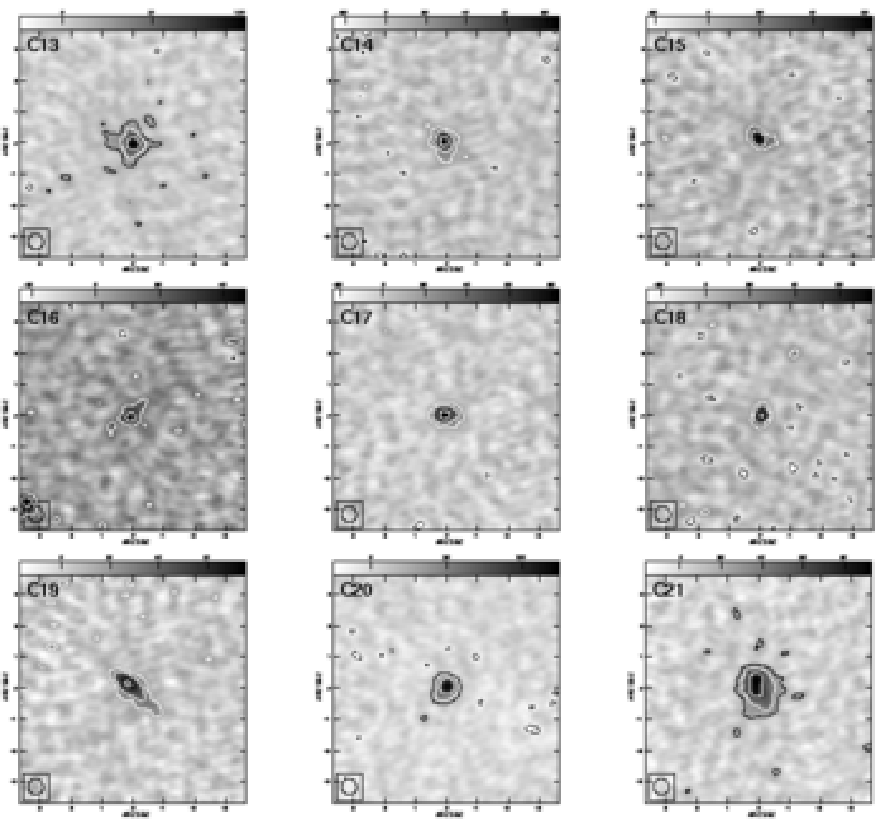}
\contcaption{}
\end{center}
\end{figure*}

\section{Observations and data reduction}
\label{obs}

\subsection{VLA}

The VLA data were first discussed in \citet{ivison02} and we give only
a brief summary here. The standard VLA wide-field `pseudo continuum'
mode was used, two separate frequency bands near 1400\,MHz (both
containing right and left circular polarisations, RCP and LCP) split
into seven 3.125-MHz-wide channels each. The correlator integration
time was 5\,s. The full primary beam was mapped, with {\sc imagr}, and
the data self-calibrated using a model of the sky (including bright
sources in the primary beam sidelobes). The data from the several days
of observations were then combined with {\sc stuffr} and produced
final maps with an r.m.s.\ noise of approximately
5\,$\mu$Jy\,beam$^{-1}$. In order to simplify later mapping of the
combined MERLIN/VLA dataset, the VLA data had all {\sc clean}
components which lay further than 6\,arcmin from the phase centre
subtracted using {\sc uvsub}.  In this way, mapping with the very
large combined dataset could be confined to the central regions of the
primary beam, the sidelobe response from brighter, more distant
sources already having been removed.

\subsection{MERLIN}

The MERLIN data were observed over the period 2005 March to June using
all seven telescopes, including the 76-m Lovell telescope which
approximately doubles the sensitivity. A single 16-MHz band with a
central frequency of 1408\,MHz was split by the correlator into 32
separate 0.5-MHz wide channels with data produced every 4\,s. The
receivers were sensitive to both RCP and LCP, but a lack of correlator
capacity does not allow the cross-correlation of the RCP and LCP data
necessary for polarisation imaging. Phase referencing was accomplished
using the nearby source, J1058+5628, with approximately 6\,min spent
observing the target and 2\,min on the calibrator. In total, the
Lockman Hole was observed for $\sim$290\,h; this corresponds to an
expected image sensitivity of $\sim$6\,$\mu$Jy\,beam$^{-1}$.  Absolute
flux calibration was based on observations of 3C\,286 and bandpass
calibration with either OQ\,208 or DA\,193.

Initial flux and bandpass calibration took place in the usual manner
with subsequent processing undertaken using the National Radio
Astronomy Observatory's {\sc aips} package. Gain solutions, first in
phase and then in amplitude and phase were derived from the phase
calibrator data and interpolated onto the target. The calibrated data
were subsequently combined into a single dataset using {\sc
stuffr}. As with \citet{muxlow05}, four of the outer channels were not
included when mapping as these reduced the image sensitivity. At this
stage it is usual to scale the data weights to reflect the nominal
sensitivity of each telescope. In preference to this, we have used
{\sc fixwt} to set the weights based on the scatter in the data; this
causes the weights to reflect the relative sensitivity of each
telescope, additionally as a function of time (caused by, for example,
the changing antenna elevation).

In preparation for combining these data with those from the VLA, maps
were made of sources identified with the VLA using {\sc imagr} with
natural weighting. Those sources which were detected and were located
further than 6\,arcmin from the phase centre had their {\sc clean}
components subtracted from the data, as with the VLA. Due to the
different primary beam response of baselines that include the Lovell
telescope, the MERLIN data were split in two, separating non-Lovell
baselines from those that include the Lovell. Each subarray was then
mapped separately and the distant sources subtracted from the data. A
complete MERLIN dataset was then reconstituted by combining the two
datasets with {\sc dbcon}.

Self-calibration of the data was attempted prior to component
subtraction. Unfortunately, due to the lack of sufficiently bright
sources, it proved impossible to find enough solutions with sufficient
signal-to-noise over the short timescales that were likely to
characterise the residual phase errors. However, the phase and
amplitude solutions are generally smooth functions of time (the
observations having been made in good weather and close to solar
minimum) and the residual phase variations are likely to be small. At
lower elevations the relative phase varies more rapidly due to the
greater differential path length, but these data have lower weight and
thus contribute less to the images. We also note that the angle
between the calibrator and target field is small, 1\fdg2.

\subsection{Bandwidth and time smearing}
\label{smearsec}

A problem inherent in imaging wide fields with radio interferometers
is chromatic aberration caused by the finite width of the frequency
channels. This results in a source being stretched along the vector
pointing from the source to the phase centre (i.e.\ radially) with a
magnitude proportional to a factor equal to the distance from the
phase centre multiplied by the fractional bandwidth. For our data, the
effect has been minimised by splitting the bandwidth into narrow
channels and by restricting our sample to sources within 6\,arcmin of
the phase centre.

The apparent size of a smeared source is a convolution of the radial
smearing factor with the telescope beam; therefore smeared sources
appear to be much bigger when observed with the VLA than with MERLIN,
even if the smearing factor is the same. In order to remove the
remaining VLA smearing we have replaced the VLA data corresponding to
each source with a de-smeared version. This is possible because at the
resolution of the VLA all our sources have very simple structures and
can be accurately modelled with single Gaussian components. These can
be subtracted from the data and replaced with new model components
that have been corrected for the appropriate amount of smearing.

The Gaussian model fitting is performed on the VLA maps using the task
{\sc jmfit}. This has a parameter ({\sc bwsmear}) which, when set to
the fractional bandwidth of the observations, corrects the fits for
bandwidth smearing and deconvolves these with the smeared beam. This
gives the source size that the VLA would have measured in the absence
of smearing. With the fit parameters fixed at the values found in the
first run, {\sc jmfit} is run again, this time without the smearing
correction. This causes the deconvolution to be performed with the
(unsmeared) {\sc clean} beam and thus returns the size of the source
as contained in the data i.e.\ the smeared size. The smeared and
unsmeared model components are then subtracted from and added to the
data respectively, using {\sc uvsub}.

Averaging visibility data in the time domain causes a similar
distortion to bandwidth smearing, but in the opposite direction
i.e.\ tangentially. Sources have visibility functions which vary more
rapidly with time at increasing distance from the phase centre and so
averaging restrictions are accordingly tighter.  With similar
correlator averaging times, the MERLIN data will have the more
restricted field of view of the two arrays. However, for our 6-arcmin
field of view we find that the effect of time smearing is negligible
for the VLA and will reduce the visibility of a MERLIN baseline by
only 1~per~cent \citep*{thompson86}. The effect in the combined maps
is also negligible given that the restoring beam is twice that of
MERLIN alone.

\subsection{Combining the MERLIN and VLA data}

Combining data from different sized arrays is a reasonably standard
procedure in radio astronomy, the resulting dataset containing a
greater range of projected baseline spacings and hence giving
simultaneous sensitivity to a wider range of source sizes. This is
perhaps most commonly associated with the different array
configurations of the VLA, but the Australian Compact Telescope Array
(ATCA) and the Westerbork Synthesis Radio Telescope (WSRT) also
routinely combine data from separate configurations. The technique has
also been used to combine data from completely different instruments,
to produce hybrid MERLIN/VLA and MERLIN/European VLBI Network (EVN)
datasets, for example.

Whilst the combination can be accomplished in the image plane, it is
instead usually done in the Fourier plane.  This is because the
improved ($u,v$) coverage is better able to constrain deconvolution
algorithms such as CLEAN and which should therefore produce more
reliable images. Having access to the visibilities has a further
advantage in that the Cotton-Schwab {\sc clean} algorithm
\citep{schwab84} can be used (as implemented in {\sc imagr}).  This is
superior to image-plane deconvolution (such as is offered by {\sc
apcln} in {\sc aips}) for a variety of reasons, including more
accurate subtraction of the {\sc clean} components and the ability to
map multiple fields simultaneously.

Whilst relatively straightforward with the single-channel datasets
that are used in the majority of radio continuum mapping, the
situation is more complicated when one is trying to image wide-field
data containing multiple channels. Within {\sc aips}, the multiple
channels are grouped together in so-called IFs (the {\sc aips} name
for a frequency band) with the ($u,v$) coordinate of a visibility in
a given channel calculable from the reference frequency of the IF and
the channel width, information which is available from the header of
the dataset.  As these numbers are generally different for a VLA and
MERLIN multi-channel dataset, it is impossible to simply combine each
by merging the IFs (as is usually done using {\sc dbcon}); calculating
the correct ($u,v$) cooordinate of a visibility would be
impossible. The problem is discussed by \citet{muxlow05} who instead
chose to combine the Dirty Images and deconvolve using {\sc apcln}.

It is, however, possible to combine the data in the Fourier plane, by
converting a multi-channel dataset into one with only a single
channel. This is achieved by extracting each channel from its IF using
{\sc uvcop}, and then re-combining with {\sc dbcon}. During this last
stage the ($u,v$) coordinates of each channel's data are set to their
correct, absolute, values and the resultant dataset has the same
structure as one with a single channel. After performing this
separately on the two IFs of the VLA and the single IF of MERLIN, the
three datasets can be further combined into one. The technique is
conceptually simple and easily implemented in {\sc aips}. No
re-weighting of data was done in {\sc dbcon} as the weights of the
MERLIN and VLA datasets were very similar.

\subsection{Mapping the combined data}

The combined MERLIN and VLA data were mapped using {\sc imagr}. We
used a robust weighting scheme, combining {\sc robust = 0} with a
Gaussian taper in the $(u,v)$ plane. This produced a roughly circular
beam ($521 \times 481$\,mas$^2$ at a position angle of 23\fdg6); this
corresponds to an improvement in resolution by a factor of nearly
three relative to the VLA-only maps (or an order of magnitude in terms
of beam area). Each map comprised 1024 $\times$ 1024 50-mas pixels and
after deconvolution had an r.m.s.\ noise of
$\sim$6.0\,$\mu$Jy\,beam$^{-1}$. We note that the quoted sensitivity
of the combined MERLIN/VLA maps of \citet{muxlow05} was better by a
factor of two compared to the single-array maps. Despite this
significant improvement, no new sources were found within the central
3-arcmin field.

We also tried making maps with a smaller beam, approximately equal to
the MERLIN-only beam ($\sim$200\,mas). However, in order to achieve
such a beamsize, the VLA data must necessarily be down-weighted and
thus contribute less to the map. These maps looked very similar to the
MERLIN-only maps and where there is extended emission it is generally
weak, patchy and over-resolved. We note that the majority of the
sources (61~per~cent) mapped by \citet{muxlow05} were restored with a
500-mas circular beam, similar to our own.

\begin{figure}
\begin{center}
\includegraphics[scale=0.4,angle=-90]{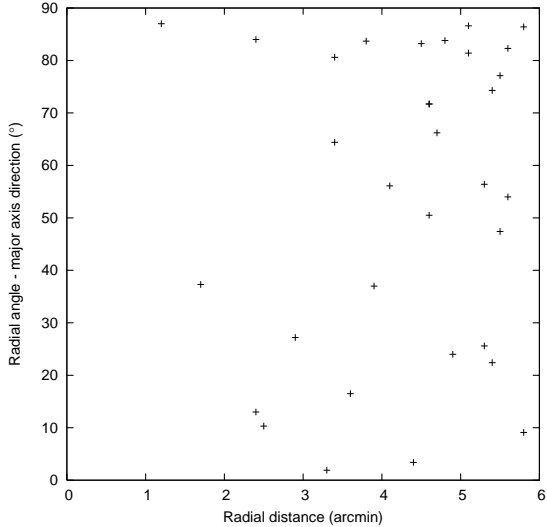}
\caption{Difference in position angle of the source radial direction
(from the phase centre) and the measured source major axis as a
function of distance from the phase centre.}
\label{smearing_test}
\end{center}
\end{figure}

MERLIN/VLA maps with an angular extent of $7.25 \times
7.25$\,arcsec$^2$ of the twelve SMGs in our sample are shown in
Fig.~\ref{smgmaps}. Contours are plotted at $-1$, 1, 2, 4, 8,
etc. times the 3\,$\sigma$ noise and greyscales are shown in units of
$\mu$Jy\,beam$^{-1}$. Maps of sources in the control sample can be
seen in Fig.~\ref{controlmaps}. Clear detections are made of all the
SMGs and control sources.

\section{Discussion}
\label{discussion}

The sources in our combined sample (SMGs and control sources) display
a variety of morphologies, but we first note that our $\sim$0.5-arcsec
beam has not resolved any of the sources into separate, multiple
components although most are extended.  In order to quantify some
properties of the sources, we have fitted a single Gaussian to each
using {\sc jmfit}. Where the sources are weak or show evidence of
multiple components the fits will be less reliable, but will still
allow us to measure the approximate source size and orientation.

Imaging such large fields with radio interferometers can lead to the
suspicion that the observed source structures are being affected by
smearing.  We are confident, however, that the steps we have taken
(restricting the sources that we image to those within a 6-arcmin
radius and correcting the VLA data for the effects of bandwidth
smearing) have resulted in our maps being unaffected by this effect to
any great degree. To test this we have computed the difference between
the measured position angle of the major axis of the Gaussian fit and
the direction of the source from the field
centre. Fig.~\ref{smearing_test} shows the difference in these angles
as a function of radial distance; the signature of smearing, i.e.\
increased clustering around zero or 90 degrees towards larger radial
distances, is not seen.

\subsection{Size of the radio-emitting region of SMGs}

The deconvolved sizes ({\sc fwhm}) of the fitted Gaussians are shown
in Table~\ref{fittab} along with a measure of the relative goodness of
fit (the sum of the square of the residuals between the model and the
data). The distribution of this un-normalised chi-square statistic has
an average of $3.7\pm0.8$ ($\times 10^{-8}$ Jy$^2$\,beam$^{-2}$) with
two significant outliers, SMG10 and C13. An examination of the maps
for these two sources shows that each is significantly resolved; the
fitted sizes for these sources should be viewed with caution. Also
included in Table~\ref{fittab} are the errors on the deconvolved sizes
determined by {\sc jmfit}.

Based on the {\sc jmfit} error analysis, all of the sources are
resolved, with deconvolved sizes greater than the associated
1-$\sigma$ error. Three sources, however, have a major axis and
uncertainty that are of similar magnitude (within a factor of two) and
a minimum size for their major axes (as returned by {\sc jmfit}) of
zero (source smaller than the beam). These three sources are the least
likely to have been resolved and they (C06, C11 and C18) are all from
the control sample.

\begin{figure}
\begin{center}
\includegraphics[scale=0.4,angle=-90]{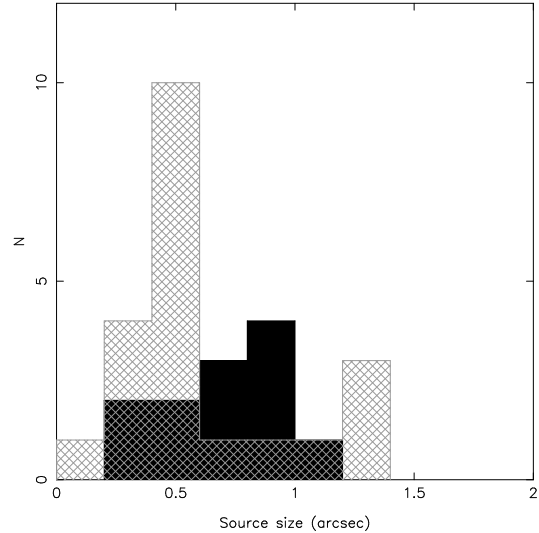}
\caption{Histogram of the source sizes (deconvolved major axis) of the
SMGs (solid) and control sample (cross-hatched) in the Lockman Hole.}
\label{sizehist}
\end{center}
\end{figure}

In Fig.~\ref{sizehist} we plot the deconvolved major axes for our
sample of SMGs. These cover the range 0.2--1.0\,arcsec, with a slight
tendency for the larger sources to be more common. The median diameter
is equal to $\sim$0.65$\pm$0.1\,arcsec.  A previous investigation into
the size of the radio emission from SMGs is reported by
\citet{chapman04}, using the combined MERLIN/VLA data from GOODS-N
that is presented in more detail by
\citet{muxlow05}. \citeauthor{chapman04} measure a median source
diameter of $0.83\pm0.14$\,arcsec, larger although not incompatible
with our own measurement.

A comparison of the two values is complicated by the different ways in
which the source sizes have been measured and the difficulty of
measuring the sizes of weak, moderately resolved sources. Whilst we
have quoted deconvolved major axes of a Gaussian fit,
\citeauthor{chapman04} give the largest extent bounded by the
3$\sigma$ contour. Gaussian fitting allows us to deconvolve the beam
from the fitted component, something that is particularly important
when the source sizes are of order the beam or smaller. Weak
extensions are not well modelled, but contribute very little to the
total flux density and would give unreliable size estimates in any
case because of the low signal to noise.

\begin{table}
\begin{center}
\caption{Deconvolved source sizes for each source. The relative
goodness of fit is indicated by the sum of squares of residuals.}
\begin{tabular}{cccc} \\ \hline
Source & Major axis & Minor axis & Sum of square of residuals \\
       &  (arcsec)  & (arcsec)   & ($\times 10^{-8}$ Jy$^2$\,beam$^{-2}$) \\ \hline
SMG01  & $0.55 \pm 0.10$ & $0.29 \pm 0.14$ & 3.4 \\
SMG02  & $0.84 \pm 0.14$ & $0.31 \pm 0.16$ & 3.6 \\
SMG03  & $0.98 \pm 0.32$ & $0.81 \pm 0.31$ & 3.0 \\
SMG04  & $0.92 \pm 0.12$ & $0.59 \pm 0.10$ & 2.9 \\
SMG05  & $0.94 \pm 0.14$ & $0.71 \pm 0.13$ & 3.6 \\
SMG06  & $0.19 \pm 0.05$ & $0.13 \pm 0.08$ & 2.9 \\
SMG07  & $0.39 \pm 0.14$ & $0.20 \pm 0.19$ & 4.3 \\
SMG08  & $0.40 \pm 0.09$ & $0.30 \pm 0.10$ & 4.4 \\
SMG09  & $0.80 \pm 0.13$ & $0.41 \pm 0.13$ & 4.2 \\
SMG10  & $0.61 \pm 0.04$ & $0.44 \pm 0.03$ & 9.7 \\
SMG11  & $0.34 \pm 0.04$ & $0.22 \pm 0.06$ & 2.0 \\
SMG12  & $0.66 \pm 0.11$ & $0.40 \pm 0.12$ & 3.0 \\
C01    & $1.16 \pm 0.20$ & $0.40 \pm 0.17$ & 2.9 \\
C02    & $0.52 \pm 0.11$ & $0.10 \pm 0.14$ & 2.9 \\
C03    & $0.49 \pm 0.13$ & $0.41 \pm 0.14$ & 5.4 \\
C04    & $0.36 \pm 0.06$ & $0.16 \pm 0.11$ & 4.3 \\
C05    & $0.46 \pm 0.09$ & $0.33 \pm 0.09$ & 3.6 \\
C06    & $0.14 \pm 0.12$ & $0.03 \pm 0.10$ & 3.3 \\
C07    & $0.83 \pm 0.16$ & $0.58 \pm 0.15$ & 3.9 \\
C08    & $0.20 \pm 0.05$ & $0.00 \pm 0.06$ & 4.8 \\
C09    & $0.43 \pm 0.15$ & $0.36 \pm 0.16$ & 3.9 \\
C010   & $0.30 \pm 0.06$ & $0.25 \pm 0.07$ & 3.9 \\
C011   & $0.25 \pm 0.19$ & $0.12 \pm 0.15$ & 4.0 \\
C012   & $0.36 \pm 0.08$ & $0.25 \pm 0.10$ & 3.1 \\
C013   & $0.66 \pm 0.08$ & $0.55 \pm 0.08$ & 8.0 \\
C014   & $0.53 \pm 0.09$ & $0.33 \pm 0.09$ & 4.4 \\
C015   & $0.50 \pm 0.11$ & $0.05 \pm 0.12$ & 2.2 \\
C016   & $1.39 \pm 0.28$ & $0.38 \pm 0.24$ & 4.2 \\
C017   & $0.52 \pm 0.07$ & $0.09 \pm 0.11$ & 2.5 \\
C018   & $0.26 \pm 0.19$ & $0.00 \pm 0.10$ & 2.6 \\
C019   & $1.29 \pm 0.14$ & $0.37 \pm 0.10$ & 5.0 \\
C020   & $0.36 \pm 0.07$ & $0.19 \pm 0.14$ & 4.5 \\
C021   & $1.09 \pm 0.08$ & $0.76 \pm 0.07$ & 4.6 \\ \hline
\end{tabular}
\label{fittab}
\end{center}
\end{table}

Measuring the sizes of SMGs is crucial to understanding their
formation. In the local universe, ULIRGs seem to be analogues of the
high-redshift SMG population \citep{sanders96}. The most luminous of
these are predominantly merger events and there is evidence that the
similarly luminous high-redshift SMGs are also commonly undergoing
mergers; physical associations are directly inferred from spectroscopy
and multi-wavelength imaging
\citep{ivison98,smail98,ivison00,conselice03,smail04,swinbank04}
whilst the double-peaked profiles often seen in mm-wave spectroscopy
of the vast reservoirs of molecular gas
\citep{frayer98,frayer99,neri03,greve05} are suggestive of orbital
motions.

\begin{figure}
\begin{center}
\includegraphics[scale=0.4,angle=-90]{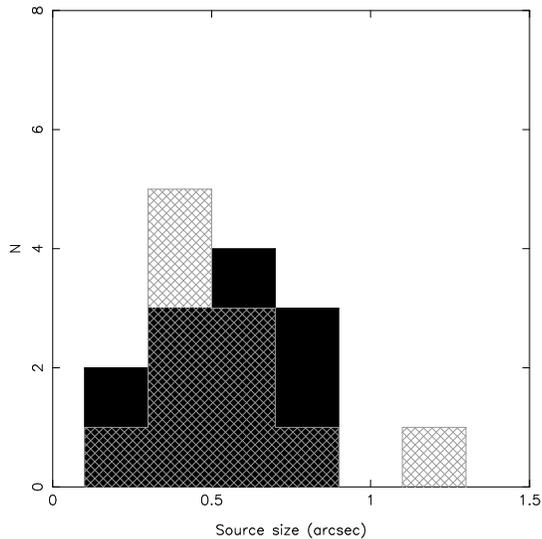}
\caption{A comparison of the sizes of SMGs as measured by our
MERLIN/VLA radio data (solid) and by \citet{tacconi06} in the mm
waveband (cross-hatched).}
\label{avsizehist}
\end{center}
\end{figure}

\citet{tacconi06} present observations at 1 and 3\,mm of SMGs with the
IRAM Plateau de Bure interferometer which at the shortest wavelength
have a beamsize ($\sim$0.6\,arcsec) that is similar to that of our
MERLIN/VLA maps.  They find that the SMGs (as seen in CO and/or
mm-wave continuum emission) are generally compact; their sample of ten
sources has a median size $\le$0.5\,arcsec ($\pm$0.1\,arcsec). The
sizes quoted by \citeauthor{tacconi06} are `angular averaged' and to
facilitate a proper comparison with our data we plot the average of
the major and minor axes of our SMGs in Fig.~\ref{avsizehist} along
with the \citeauthor{tacconi06} results.  The two distributions appear
similar, in both the position of the peaks and the range of sizes
covered.  A Kolmogorov-Smirnov (K-S) test \citep{press92} gives a
probability of 84~per~cent that the differences between the two could
arise by chance and thus strongly favours the null hypothesis that the
two distributions are the same. This mirrors what is seen at much
lower redshift, where the radio emission \citep{condon91} and
molecular gas \citep{downes98} have similar sizes.

Converting to linear distances for the SMGs with redshifts\footnote{We
assume a flat `concordance' cosmology with $\Omega_{\rm m} = 0.27$,
$\Omega_{\Lambda} = 0.73$ and $H_0 = 71$\,km\,s$^{-1}$\,Mpc$^{-1}$
\citep{spergel03}.} we find a range 1.2--7.7\,kpc with a median of
4.9\,kpc; \citet{tacconi06} measure a smaller, but similar, linear
extent of $<$4\,kpc. Given that the size of the central starburst in
local ULIRGs is $\la$1\,kpc, the high-redshift starbursts appear to be
in general significantly larger, although the smallest (SMG06) is of
comparable size to those seen at low redshift.

\begin{figure}
\begin{center}
\includegraphics[scale=0.4]{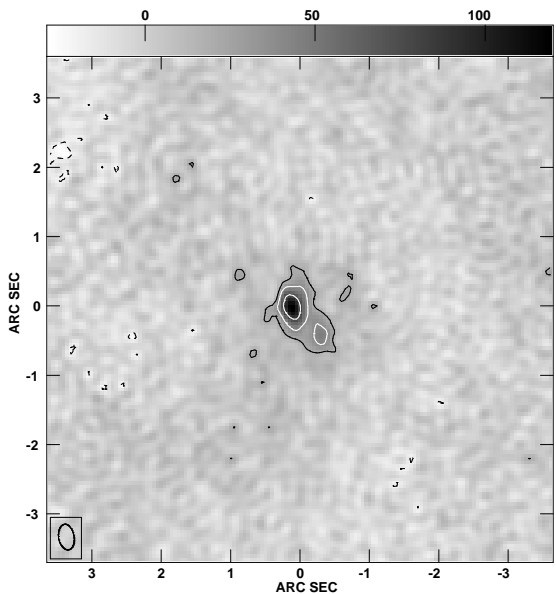}
\caption{MERLIN-only map of SMG09 (LH\,1200.008). The beam is shown in
the bottom-left corner and has dimensions of $373 \times 223$\,mas$^2$
at a position angle of $9\fdg7$.  The greyscale has units of
$\mu$Jy\,beam$^{-1}$ and contours are drawn at $-1$, 1, 2, 4, 8,
etc.\ times the 3\,$\sigma$ noise (3 $\times$
8.5\,$\mu$Jy\,beam$^{-1}$).}
\label{1200.008fig}
\end{center}
\end{figure}

\citet{ivison07} have explored the fraction of SMGs that contain
multiple radio counterparts and find evidence that this is more common
for the brighter sources, with separations between 2 and
6\,arcsec\footnote{SMG02 (Lock\,850.04) has two radio components close
to the submm position and although only one of these is included in
our sample, the weaker of the two ($<$50$\mu$Jy) is detected in the
MERLIN/VLA map and is visible as the 3$\sigma$ contour north-northeast
of the stronger detection.}. They are unable to probe below 2\,arcsec
due to the resolution of the VLA-only data, but despite the smaller
beam of our combined MERLIN/VLA maps we see no examples of multiple
sources with separations $<$2\,arcsec. We can increase the resolution
still further, at the expense of accurate imaging of extended
emission, by examining the MERLIN-only maps. One of these, SMG10,
reveals that a prominent extension in the MERLIN/VLA map contains a
relatively high signal-to-noise ($\sim$6) component
(Fig.~\ref{1200.008fig}). A two-Gaussian fit in {\sc jmfit} gives a
separation of 0.6\,arcsec and a flux density ratio of $\sim$5:1. At the
redshift of 1.212 this angular separation corresponds to a linear
distance of $\sim$5\,kpc. The nature of this second component is
unknown and may be a radio jet, but it may also be a small-separation
merger.

\subsection{SMG sizes relative to the $\mu$Jy radio population}

\begin{figure}
\begin{center}
\includegraphics[scale=0.4,angle=-90]{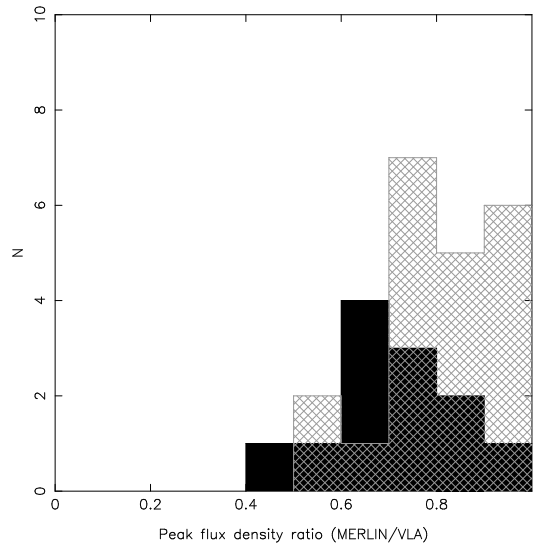}
\caption{Histogram of the peak flux density ratio (MERLIN/VLA) of the
SMGs (solid) and control sample (cross-hatched) in the Lockman Hole.}
\label{peakhist}
\end{center}
\end{figure}

Also plotted in Fig.~\ref{sizehist} are the major axes of the fits to
the control sample, i.e.\ those $\mu$Jy radio sources which are not
luminous in the submm waveband. The range of angular sizes for both
samples is similar, covering approximately 0.1--1.4\,arcsec although
the median size of the control sample is formally smaller than that of
the SMGs, 0.5\,arcsec compared to 0.65\,arcsec (with 1\,$\sigma$
errors of about 0.1\,arcsec).  Another way of assessing the extension
or otherwise of the two source populations is to compare the peak flux
densities in the MERLIN-only and combined MERLIN/VLA maps. These
ratios are plotted in Fig.~\ref{peakhist} and a similar picture
emerges; the control source distribution contains sources which are,
in general, more compact. A K-S analysis gives low probabilities that
the distributions are the same (30 and 17~per~cent for the size and
peak flux density plots, respectively).

\begin{figure}
\begin{center}
\includegraphics[scale=0.4,angle=-90]{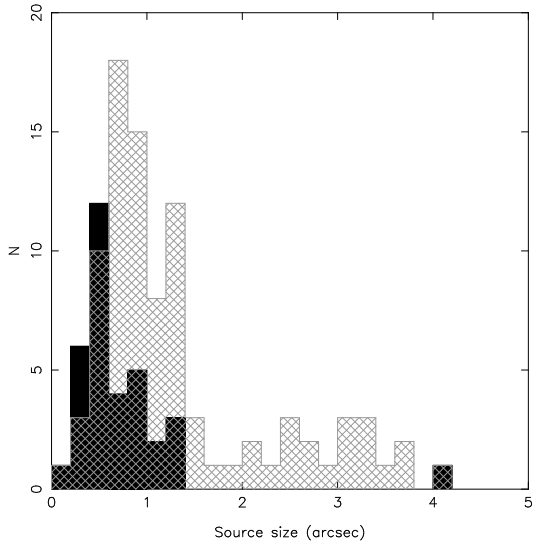}
\caption{Histogram of the source sizes of our Lockman Hole sample as
well as the faint radio population in the GOODS-N field of
\citet{muxlow05}. For the Lockman Hole (solid area) the sizes refer to
the deconvolved major axis of a Gaussian fit whilst for GOODS-N
(cross-hatched) it is the `largest angular size'.}
\label{me_tom_hist}
\end{center}
\end{figure}

We plot the sizes of all our sources in Fig.~\ref{me_tom_hist} and
include the bright source that was excluded from our control sample (a
large, $\sim$4\,arcsec, source which contains a radio core and
two-sided jet). The distribution peaks strongly at angular extents in
the range 0.4--0.6\,arcsec. This is in apparent disagreement with
\citet{muxlow05} and \citet{chapman04} who find that the average
source size of $\mu$Jy radio sources is $\sim$1\,arcsec and that there
is a prominent tail extending to $\sim$4\,arcsec. The
\citeauthor{muxlow05} data are also plotted in Fig.~\ref{me_tom_hist}
so that the two samples can be easily compared; the two distributions
are very different\footnote{Our histogram of the \citeauthor{muxlow05}
data is not identical to that in the original publication, probably
due to slight differences in the way the data are binned.} and they
can formally be rejected as representing the same population with a
K-S probability in excess of 99~per~cent.

The distributions in Fig.~\ref{me_tom_hist} differ in two
ways. Firstly, the Lockman Hole data appear to consist of only a
single source population, this being distributed approximately
normally. The GOODS-N distribution is similar for the smaller sources,
but peaks at a larger source size and has a significant tail out to
angular extents of $\sim$4\,arcsec. The absence of this tail in our
sample (save for a single source) is particularly striking. Even
taking into account possible errors in our model fitting (relatively
weak extensions are not well modelled), we can be certain that only
one source in our 12-arcmin-diameter field is larger than 2\,arcsec.

The explanation for the difference in source size distributions in
Fig.~\ref{me_tom_hist}, especially with regard to the tail, is
probably a combination of the larger sample and lower flux limit of
\citeauthor{muxlow05} (40\,$\mu$Jy compared to our 50\,$\mu$Jy) and
cosmic variance. Our sources may also be smaller, particularly for
sources in the population which peaks at around 0.5--1\,arcsec,
because of differences in the way the sizes were measured and also
because we have attempted to remove the bandwidth smearing from our
VLA data, something that was not done for the GOODS-N data.

\section{Conclusions}
\label{conclusions}

We have presented sensitive maps (6\,$\mu$Jy\,beam$^{-1}$ r.m.s.) of
12 SMGs in the Lockman Hole made from combined MERLIN and VLA
data. This is the first time that wide-field (multi-channel) data from
MERLIN and the VLA have been combined in the $(u,v)$ plane. The
procedure to do this is straightforward, practical with modern desktop
computing facilities and can be accomplished with existing software
tools, i.e.\ {\sc aips}.

From the radio morphology alone it has generally not been possible to
determine unambiguously whether a source contains an AGN, except in
one case which has a prominent jet (C19). In the local Universe, radio
emission from most starburst galaxies is known to be concentrated in
their galactic nuclei, so source compactness is no guarantee of an
AGN. Indeed, even higher resolution -- very long baseline
interferometry -- is required to improve our diagnostic ability, via
sensitivity to more compact jets and via the ability to constrain the
brightness temperature of unresolved emission to a regime unique to
AGN.

Our maps have resolved the vast majority of the SMGs as well as a
complementary sample of radio sources of similar flux density. A
comparison of the properties of these two populations shows that the
size distributions overlap considerably, but the SMGs are in general
larger than the control sample (six of which harbour AGN, being
luminous X-ray sources -- see below). Even if some of our
submm-luminous sources also contain significant emission from AGN,
this suggests that the starburst component in high-redshift SMGs is
extended and not confined to the nucleus.

This conclusion is strengthened by the measurement of similar sizes
for the radio-emitting regions of SMGs and their molecular gas
reservoirs, suggesting that AGN are not significantly contributing to
the measured size of the radio emission. However, most high-redshift
SMGs have much larger sizes (1.2--7.7\,kpc) than local ULIRGs
($\le$1\,kpc), possibly the result of witnessing them at a less
advanced stage of the merger process.

Despite our high resolution, as compared with the VLA, we find only
one source that is resolved into multiple, distinct
components. Determining the true nature of this source (and SMG06) --
merger-induced starburst or AGN plus jet? -- will require follow-up
observations, but small separation ($<$2\,arcsec) multiple
counterparts to SMGs appear to be rare.

X-ray emission is a good indicator of AGN activity. Based on
counterparts in the {\em XMM-Newton} catalogue of \citet{brunner07},
we find a lower detection rate of X-ray luminous AGN amongst the
submm-luminous sources than amongst the control sample (Poissonian
probability, 89 per cent). Six of the control sources have a likely
X-ray counterpart (C05, C10, C12, C13, C20 and C21) versus only one of
the submm galaxies (SMG06). SMG06 is the hardest of the seven X-ray
detections, consistent with enhanced X-ray absorption in a
dust-obscured starburst \citep{page04}. It is also the most compact
radio emitter amongst the SMGs, with a flat radio spectrum
\citep{ivison07}.

$e$-MERLIN and the EVLA will soon be available, the broad-banded and
hence much more sensitive versions of the current arrays. This will
enable the $\mu$Jy radio population to be imaged at high resolution
over larger areas and at increasingly faint flux limits. Also to come
is the eSMA, a collaboration between the JCMT, the CSO and the
telescopes of the Submillimeter Array. The resulting instrument will
have a resolution at 345\,GHz of 300\,mas, extremely complementary to
the angular scales being probed by the combined MERLIN and VLA maps
presented in this paper. This will allow a direct comparison between
the submm and radio morphologies of SMGs and we can thus examine the
radio/far-IR correlation at unprecedented resolution.

\section*{Acknowledgments}

Many thanks to Tom Muxlow, Ian Smail and James Dunlop for their
advice. Based on observations made with MERLIN, a National Facility
operated by the University of Manchester at Jodrell Bank Observatory
on behalf of the Science and Technology Facilities Council. The
National Radio Astronomy Observatory is a facility of the National
Science Foundation operated under cooperative agreement by Associated
Universities, Inc.

\bibliographystyle{mnras}
\bibliography{deep}

\begin{thebibliography}{}

\bibitem[\protect\citeauthoryear{{Alexander} et~al.}{{Alexander}
  et~al.}{2003}]{alexander03}
{Alexander} D.~M. et~al., 2003, \aj, 125, 383

\bibitem[\protect\citeauthoryear{{Alexander} et~al.}{{Alexander}
  et~al.}{2005}]{alexander05}
{Alexander} D.~M., {Bauer} F.~E., {Chapman} S.~C., {Smail} I., {Blain} A.~W.,
  {Brandt} W.~N.,  {Ivison} R.~J., 2005, \apj, 632, 736

\bibitem[\protect\citeauthoryear{{Biggs} \& {Ivison}}{{Biggs} \&
  {Ivison}}{2006}]{biggs06}
{Biggs} A.~D.,  {Ivison} R.~J., 2006, \mnras, 371, 963

\bibitem[\protect\citeauthoryear{{Brunner} et~al.}{{Brunner}
  et~al.}{2007}]{brunner07}
{Brunner} H., {Cappelluti} N., {Hasinger} G., {Barcons} X., {Fabian} A.~C.,
  {Mainieri} V.,  {Szokoly} G., 2007, \aap, 711, astro-ph/0711.4822

\bibitem[\protect\citeauthoryear{{Chapman} et~al.}{{Chapman}
  et~al.}{2003}]{chapman03}
{Chapman} S.~C., {Blain} A.~W., {Ivison} R.~J.,  {Smail} I.~R., 2003, \nat,
  422, 695

\bibitem[\protect\citeauthoryear{{Chapman} et~al.}{{Chapman}
  et~al.}{2005}]{chapman05}
{Chapman} S.~C., {Blain} A.~W., {Smail} I.,  {Ivison} R.~J., 2005, \apj, 622,
  772

\bibitem[\protect\citeauthoryear{{Chapman} et~al.}{{Chapman}
  et~al.}{2004}]{chapman04}
{Chapman} S.~C., {Smail} I., {Windhorst} R., {Muxlow} T.,  {Ivison} R.~J.,
  2004, \apj, 611, 732

\bibitem[\protect\citeauthoryear{{Ciliegi} et~al.}{{Ciliegi}
  et~al.}{2000}]{ciliegi00}
{Ciliegi} P., {Zamorani} G., {Gruppioni} C., {Hasinger} G., {Lehmann} I.,
  {Wilson} G., 2000, in {Plionis} M.,  {Georgantopoulos} I., eds, Large Scale
  Structure in the X-ray Universe, Proceedings of the 20-22 September 1999
  Workshop, Santorini, Greece.
\newblock Atlantisciences, Paris, p. 347

\bibitem[\protect\citeauthoryear{{Condon} et~al.}{{Condon}
  et~al.}{1991}]{condon91}
{Condon} J.~J., {Huang} Z.-P., {Yin} Q.~F.,  {Thuan} T.~X., 1991, \apj, 378, 65

\bibitem[\protect\citeauthoryear{{Conselice}, {Chapman} \&
  {Windhorst}}{{Conselice} et~al.}{2003}]{conselice03}
{Conselice} C.~J., {Chapman} S.~C.,  {Windhorst} R.~A., 2003, \apjl, 596, L5

\bibitem[\protect\citeauthoryear{{Coppin} et~al.}{{Coppin}
  et~al.}{2006}]{coppin06}
{Coppin} K. et~al., 2006, \mnras, 372, 1621

\bibitem[\protect\citeauthoryear{{De Ruiter} et~al.}{{De Ruiter}
  et~al.}{1997}]{deruiter97}
{De Ruiter} H.~R. et~al., 1997, \aap, 319, 7

\bibitem[\protect\citeauthoryear{{Di Matteo}, {Springel} \& {Hernquist}}{{Di
  Matteo} et~al.}{2005}]{dimatteo05}
{Di Matteo} T., {Springel} V.,  {Hernquist} L., 2005, \nat, 433, 604

\bibitem[\protect\citeauthoryear{{Downes} \& {Solomon}}{{Downes} \&
  {Solomon}}{1998}]{downes98}
{Downes} D.,  {Solomon} P.~M., 1998, \apj, 507, 615

\bibitem[\protect\citeauthoryear{{Egami} et~al.}{{Egami}
  et~al.}{2004}]{egami04}
{Egami} E. et~al., 2004, \apjs, 154, 130

\bibitem[\protect\citeauthoryear{{Fox} et~al.}{{Fox} et~al.}{2002}]{fox02}
{Fox} M.~J. et~al., 2002, \mnras, 331, 839

\bibitem[\protect\citeauthoryear{{Frayer} et~al.}{{Frayer}
  et~al.}{1999}]{frayer99}
{Frayer} D.~T. et~al., 1999, \apjl, 514, L13

\bibitem[\protect\citeauthoryear{{Frayer} et~al.}{{Frayer}
  et~al.}{1998}]{frayer98}
{Frayer} D.~T., {Ivison} R.~J., {Scoville} N.~Z., {Yun} M., {Evans} A.~S.,
  {Smail} I., {Blain} A.~W.,  {Kneib} J.-P., 1998, \apjl, 506, L7

\bibitem[\protect\citeauthoryear{{Garrett} et~al.}{{Garrett}
  et~al.}{2001}]{garrett01}
{Garrett} M.~A. et~al., 2001, \aap, 366, L5

\bibitem[\protect\citeauthoryear{{Greve} et~al.}{{Greve}
  et~al.}{2005}]{greve05}
{Greve} T.~R. et~al., 2005, \mnras, 359, 1165

\bibitem[\protect\citeauthoryear{{Greve} et~al.}{{Greve}
  et~al.}{2004}]{greve04}
{Greve} T.~R., {Ivison} R.~J., {Bertoldi} F., {Stevens} J.~A., {Dunlop} J.~S.,
  {Lutz} D.,  {Carilli} C.~L., 2004, \mnras, 354, 779

\bibitem[\protect\citeauthoryear{{H{\"a}ring} \& {Rix}}{{H{\"a}ring} \&
  {Rix}}{2004}]{haring04}
{H{\"a}ring} N.,  {Rix} H.-W., 2004, \apjl, 604, L89

\bibitem[\protect\citeauthoryear{{Hasinger} et~al.}{{Hasinger}
  et~al.}{2001}]{hasinger01}
{Hasinger} G. et~al., 2001, \aap, 365, L45

\bibitem[\protect\citeauthoryear{{Holland} et~al.}{{Holland}
  et~al.}{1999}]{holland99}
{Holland} W.~S. et~al., 1999, \mnras, 303, 659

\bibitem[\protect\citeauthoryear{{Huang} et~al.}{{Huang}
  et~al.}{2004}]{huang04}
{Huang} J.-S. et~al., 2004, \apjs, 154, 44

\bibitem[\protect\citeauthoryear{{Hughes} et~al.}{{Hughes}
  et~al.}{1998}]{hughes98}
{Hughes} D.~H. et~al., 1998, \nat, 394, 241

\bibitem[\protect\citeauthoryear{{Ivison} et~al.}{{Ivison}
  et~al.}{2007}]{ivison07}
{Ivison} R.~J. et~al., 2007, \mnras, 380, 199

\bibitem[\protect\citeauthoryear{{Ivison} et~al.}{{Ivison}
  et~al.}{2002}]{ivison02}
{Ivison} R.~J. et~al., 2002, \mnras, 337, 1

\bibitem[\protect\citeauthoryear{{Ivison} et~al.}{{Ivison}
  et~al.}{2000}]{ivison00}
{Ivison} R.~J., {Smail} I., {Barger} A.~J., {Kneib} J.-P., {Blain} A.~W.,
  {Owen} F.~N., {Kerr} T.~H.,  {Cowie} L.~L., 2000, \mnras, 315, 209

\bibitem[\protect\citeauthoryear{{Ivison} et~al.}{{Ivison}
  et~al.}{2005}]{ivison05}
{Ivison} R.~J. et~al., 2005, \mnras, 364, 1025

\bibitem[\protect\citeauthoryear{{Ivison} et~al.}{{Ivison}
  et~al.}{1998}]{ivison98}
{Ivison} R.~J., {Smail} I., {Le Borgne} J.-F., {Blain} A.~W., {Kneib} J.-P.,
  {Bezecourt} J., {Kerr} T.~H.,  {Davies} J.~K., 1998, \mnras, 298, 583

\bibitem[\protect\citeauthoryear{{Laurent} et~al.}{{Laurent}
  et~al.}{2005}]{laurent05}
{Laurent} G.~T. et~al., 2005, \apj, 623, 742

\bibitem[\protect\citeauthoryear{{Laurent} et~al.}{{Laurent}
  et~al.}{2006}]{laurent06}
{Laurent} G.~T. et~al., 2006, \apj, 643, 38

\bibitem[\protect\citeauthoryear{{Lockman}, {Jahoda} \& {McCammon}}{{Lockman}
  et~al.}{1986}]{lockman86}
{Lockman} F.~J., {Jahoda} K.,  {McCammon} D., 1986, \apj, 302, 432

\bibitem[\protect\citeauthoryear{{Lutz} et~al.}{{Lutz} et~al.}{2005}]{lutz05}
{Lutz} D., {Valiante} E., {Sturm} E., {Genzel} R., {Tacconi} L.~J., {Lehnert}
  M.~D., {Sternberg} A.,  {Baker} A.~J., 2005, \apjl, 625, L83

\bibitem[\protect\citeauthoryear{{Magorrian} et~al.}{{Magorrian}
  et~al.}{1998}]{magorrian98}
{Magorrian} J. et~al., 1998, \aj, 115, 2285

\bibitem[\protect\citeauthoryear{{Mainieri} et~al.}{{Mainieri}
  et~al.}{2002}]{mainieri02}
{Mainieri} V., {Bergeron} J., {Hasinger} G., {Lehmann} I., {Rosati} P.,
  {Schmidt} M., {Szokoly} G.,  {Della Ceca} R., 2002, \aap, 393, 425

\bibitem[\protect\citeauthoryear{{Marconi} \& {Hunt}}{{Marconi} \&
  {Hunt}}{2003}]{marconi03}
{Marconi} A.,  {Hunt} L.~K., 2003, \apjl, 589, L21

\bibitem[\protect\citeauthoryear{{Men{\'e}ndez-Delmestre}
  et~al.}{{Men{\'e}ndez-Delmestre} et~al.}{2007}]{menendez07}
{Men{\'e}ndez-Delmestre} K. et~al., 2007, \apjl, 655, L65

\bibitem[\protect\citeauthoryear{{Mortier} et~al.}{{Mortier}
  et~al.}{2005}]{mortier05}
{Mortier} A.~M.~J. et~al., 2005, \mnras, 363, 563

\bibitem[\protect\citeauthoryear{{Muxlow} et~al.}{{Muxlow}
  et~al.}{2005}]{muxlow05}
{Muxlow} T.~W.~B. et~al., 2005, \mnras, 358, 1159

\bibitem[\protect\citeauthoryear{{Neri} et~al.}{{Neri} et~al.}{2003}]{neri03}
{Neri} R. et~al., 2003, \apjl, 597, L113

\bibitem[\protect\citeauthoryear{{Page} et~al.}{{Page} et~al.}{2004}]{page04}
{Page} M.~J., {Stevens} J.~A., {Ivison} R.~J.,  {Carrera} F.~J., 2004, \apjl,
  611, L85

\bibitem[\protect\citeauthoryear{{Press} et~al.}{{Press}
  et~al.}{1992}]{press92}
{Press} W.~H., {Teukolsky} S.~A., {Vetterling} W.~T.,  {Flannery} B.~P., 1992,
  {Numerical recipes in FORTRAN}.
\newblock Cambridge Univ. Press, Cambridge

\bibitem[\protect\citeauthoryear{{Sanders} \& {Mirabel}}{{Sanders} \&
  {Mirabel}}{1996}]{sanders96}
{Sanders} D.~B.,  {Mirabel} I.~F., 1996, \araa, 34, 749

\bibitem[\protect\citeauthoryear{{Schwab}}{{Schwab}}{1984}]{schwab84}
{Schwab} F.~R., 1984, \aj, 89, 1076

\bibitem[\protect\citeauthoryear{{Scott} et~al.}{{Scott}
  et~al.}{2002}]{scott02}
{Scott} S.~E. et~al., 2002, \mnras, 331, 817

\bibitem[\protect\citeauthoryear{{Silk} \& {Rees}}{{Silk} \&
  {Rees}}{1998}]{silk98}
{Silk} J.,  {Rees} M.~J., 1998, \aap, 331, L1

\bibitem[\protect\citeauthoryear{{Smail} et~al.}{{Smail}
  et~al.}{2004}]{smail04}
{Smail} I., {Chapman} S.~C., {Blain} A.~W.,  {Ivison} R.~J., 2004, \apj, 616,
  71

\bibitem[\protect\citeauthoryear{{Smail}, {Ivison} \& {Blain}}{{Smail}
  et~al.}{1997}]{smail97}
{Smail} I., {Ivison} R.~J.,  {Blain} A.~W., 1997, \apjl, 490, L5

\bibitem[\protect\citeauthoryear{{Smail} et~al.}{{Smail}
  et~al.}{1998}]{smail98}
{Smail} I., {Ivison} R.~J., {Blain} A.~W.,  {Kneib} J.-P., 1998, \apjl, 507,
  L21

\bibitem[\protect\citeauthoryear{{Spergel} et~al.}{{Spergel}
  et~al.}{2003}]{spergel03}
{Spergel} D.~N. et~al., 2003, \apjs, 148, 175

\bibitem[\protect\citeauthoryear{{Strom}}{{Strom}}{2004}]{strom04}
{Strom} R., 2004, in {Bachiller} R., {Colomer} F., {Desmurs} J.-F.,  {de
  Vicente} P., eds, European VLBI Network on New Developments in VLBI Science
  and Technology, p. 271

\bibitem[\protect\citeauthoryear{{Swinbank} et~al.}{{Swinbank}
  et~al.}{2004}]{swinbank04}
{Swinbank} A.~M., {Smail} I., {Chapman} S.~C., {Blain} A.~W., {Ivison} R.~J.,
  {Keel} W.~C., 2004, \apj, 617, 64

\bibitem[\protect\citeauthoryear{{Tacconi} et~al.}{{Tacconi}
  et~al.}{2006}]{tacconi06}
{Tacconi} L.~J. et~al., 2006, \apj, 640, 228

\bibitem[\protect\citeauthoryear{{Thompson}, {Moran} \& {Swenson}}{{Thompson}
  et~al.}{1986}]{thompson86}
{Thompson} A.~R., {Moran} J.~M.,  {Swenson} G.~W., 1986, {Interferometry and
  synthesis in radio astronomy}.
\newblock Wiley-Interscience, New York

\end{thebibliography}

\bsp

\end{document}